\documentclass[12pt]{article} 
\usepackage[sectionbib]{natbib}
\usepackage{array,epsfig,fancyheadings,rotating}
\usepackage[]{hyperref}  
\usepackage{sectsty, secdot}
\sectionfont{\fontsize{12}{14pt plus.8pt minus .6pt}\selectfont}
\renewcommand{\theequation}{\thesection\arabic{equation}}
\subsectionfont{\fontsize{12}{14pt plus.8pt minus .6pt}\selectfont}

\usepackage{amsmath}
\usepackage{amssymb}
\usepackage{amsfonts}
\usepackage{multirow}
\usepackage{amsthm}
\usepackage{color,xcolor}
\usepackage{esvect}
\usepackage{threeparttable}
\usepackage[]{caption2}
\usepackage[noblocks]{authblk}
\usepackage{array}

\newtheorem{theorem}{Theorem}

\theoremstyle{definition}

\begin{document}


\renewcommand{\baselinestretch}{2}


\renewcommand{\thefootnote}{\fnsymbol{footnote}}
$\ $\par


\fontsize{12}{14pt plus.8pt minus .6pt}\selectfont \vspace{0.8pc}
\centerline{\large\bf Null-Free False Discovery Rate Control}
\vspace{2pt} \centerline{\large\bf  Using Decoy Permutations}

\vspace{1pt} \centerline{Kun He$^{1,3}$, Mengjie Li$^{2,3}$, Yan Fu$^{2,3}$\footnote{To whom correspondence should be addressed: yfu@amss.ac.cn}, Fuzhou Gong$^{2,3}$, Xiaoming Sun$^{1,3}$} 
\vspace{1pt} \centerline{\it $^1$Institute of Computing Technology, Chinese Academy of Sciences} 
\vspace{1pt} \centerline{\it $^2$Academy of Mathematics and Systems Science, Chinese Academy of Sciences} 
\vspace{1pt} \centerline{\it $^3$University of Chinese Academy of Sciences}

\vspace{.55cm} \fontsize{9}{11.5pt plus.8pt minus
.6pt}\selectfont


\begin{quotation}
\noindent {\it Abstract:} The traditional approaches to false discovery rate (FDR) control in multiple hypothesis testing are usually based on the null distribution of a test statistic. However, all types of null distributions, including the theoretical, permutation-based and empirical ones, have some inherent drawbacks. For example, the theoretical null might fail because of improper assumptions on the sample distribution. 
Here, we propose a null distribution-free approach to FDR control for multiple hypothesis testing. This approach, named \emph{target-decoy procedure}, simply builds on the ordering of tests by some statistic or score, the null distribution of which is not required to be known. Competitive decoy tests are constructed from permutations of original samples and are used to estimate the false target discoveries. We prove that this approach controls the FDR when the statistics are independent between different tests. Simulation demonstrates that it is more stable and powerful than two existing popular approaches. Evaluation is also made on a real dataset. 

\vspace{9pt}
\par
\end{quotation}\par

\def\thefigure{\arabic{figure}}
\def\thetable{\arabic{table}}

\renewcommand{\theequation}{\thesection.\arabic{equation}}

\fontsize{12}{14pt plus.8pt minus .6pt}\selectfont

\setcounter{section}{0} 
\setcounter{equation}{0} 


\section{Introduction}
\label{sec:intro}
\subsection{Traditional approaches to FDR control}
Multiple testing has become increasingly popular in the present big-data era. For example, a typical scenario of applying multiple testing in biomedical studies is to look for differentially expressed genes/proteins, from thousands of candidates, between two groups (i.e. cases and controls) of samples \citep{efron2008microarrays, diz2011multiple}. 
Currently, controlling the false discovery rate (FDR), which is defined as the expected proportion of incorrect rejections among all rejections \citep{benjamini1995controlling}, is the predominant way to do multiple testing. 
FDR control procedures aim at selecting a subset of rejected hypotheses such that the FDR is no more than a given level.

Because a \emph{p}-value is typically computed from the null distribution of a test statistic in each single test, the canonical approaches to FDR control for multiple testing at present are based on the \emph{p}-values of all tests or at least the null distribution of the test statistic.
Since \citet{benjamini1995controlling} proposed the first \emph{p}-value based sequential procedure to control the FDR (BH procedure), many FDR control approaches have been developed, e.g., \citep{benjamini2001control,sarkar2002some,storey2002direct, storey2003positive,benjamini2006adaptive,basu2017weighted}.

A key problem faced by these approaches is how to obtain the proper null distribution. Popular null distributions, including the theoretical null, permutation null and empirical null, often suffer one way or another \citep{efron2008microarrays,efron2012large}. The theoretical null, though widely used, might fail in practice for many reasons, such as improper mathematical assumptions or unobserved covariates \citep{efron2007size, efron2008microarrays}. For example, for the Student's \emph{t}-test, if the sample  distribution is not normal, the \emph{t}-value will not follow a \emph{t}-distribution and the \emph{p}-values calculated will not be uniform (0, 1) distributed for true null hypotheses. The permutation null is also widely used. There are mainly two different permutation methods, i.e., the permutation tests and the pooled permutation \citep{kerr2009comments}. The permutation tests are a class of widely used non-parametric tests to calculate \emph{p}-values, and are most useful when the information about the data distribution is insufficient. However, the statistical power of permutation tests is limited by the sample size of a test \citep{tusher2001significance}. Instead of estimating a null distribution for each test individually, the pooled permutation in multiple testing estimates an overall null distribution for all tests \citep{efron2001empirical}. However, it has been found that pooling permutation null distributions across hypotheses can produce invalid \emph{p}-values, since even true null hypotheses can have different permutation distributions \citep{kerr2009comments}. 

\par To overcome the shortcomings of the theoretical and permutation null distributions, new methods were proposed to estimate an empirical null distribution from a large number of tests \citep{efron2001empirical, efron2002empirical, efron2008microarrays,scott2010bayes}. For example, the empirical Bayes method estimates the empirical null distribution by decomposing the mixture of null and alternative distributions \citep{efron2008microarrays}. However, decomposing the mixture distribution is intrinsically a difficult problem. For example, if the empirical distribution has a strong peak, the decomposing may fail \citep{strimmer2008unified}.

Moreover, the proportion of true null hypotheses has to be estimated either explicitly or implicitly to apply these FDR control methods. If this null proportion is ignored (e.g., assumed to be one as in the original BH procedure), the power of testing would be reduced. Since \citet{storey2002direct} proposed the first approach, estimation of the null proportion has become a key component of current FDR methods to enhance the power, such as the Bayes and the empirical Bayes methods \citep{storey2003positive, storey2004strong, benjamini2006adaptive, efron2008microarrays, strimmer2008unified}.
More accurate estimation of the null ratio has been of great interest in the field \citep{langaas2005estimating,meinshausen2006estimating,markitsis2010censored,yu2017parametric}.

\subsection{Our approach to FDR control}
Here, we propose a new approach to FDR control, named \emph{target-decoy procedure}, which is free of the null distribution and the null proportion. In this approach, a target score and a number of decoy scores are calculated for each test. 
These scores are used to measure the (dis)similarities of two groups of samples, and can be popular statistics, e.g., \emph{t}-value or other scoring functions.  
The target score is calculated with regard to the original samples, while the decoy scores are calculated with regard to randomly permuted samples. Based on the target score and decoy scores, a label and a final score are calculated for each test in a competitive manner. For example, in the simplified target-decoy procedure if the target score is more significant than half of the decoy scores, the test is labelled as target and the final score is set as the target score. Otherwise, if the target score is less significant than half of the decoy scores, the test is labelled as decoy and the final score is set as the decoy score with a specific rank that is mapped symmetrically from the rank of the target score. Then the tests are sorted by their final scores and the ratio of the number (added by one) of decoy test statistics to the number of target test statistics beyond a threshold is used for FDR control. We prove that such target-decoy procedure can rigorously control the FDR when the scores are independent between tests.  

Our approach is exclusively based on the scores and labels of tests. The scoring function used is not limited to traditional \emph{p}-value or test statistics which have clear null distributions, but can be in any free forms with some symmetry property. Therefore, our approach provides great flexibility and can be potentially more powerful than traditional approaches, the performance of which largely relies on the precision of \emph{p}-values or the sample size of each test. 
Monte-Carlo simulations demonstrate that our approach effectively controls the FDR and is more powerful than two popular methods, i.e., the Bayes method \citep{storey2002direct, storey2003positive, storey2004strong} and the empirical Bayes method \citep{efron2001empirical, efron2002empirical, efron2008microarrays}. The performances of the three methods were also compared on a real dataset. Because our procedure is more straightforward and can be used with arbitrary score functions, we believe that it will have many practical applications.

The rest of the paper is organized as follows. Section \ref{sec: section 2} describes our target-decoy approach for FDR control. Section \ref{sec:Problem Formulation} discusses a general scenario of case-control study. The simplified and standard target-decoy procedures are presented in Sections \ref{sec:simplified target-decoy procedure} and \ref{sec:The target-decoy approach}, respectively. Section \ref{sec:Automatically choosing $r$} provides an adaptive version of the target-decoy procedure. Section \ref{sec:Theoretic Analysis} establishes the theoretical foundation of our approach (Proofs are given in Supplementary Material). Numerical results on independent and dependent variables are given in Section \ref{sec:Simulation Studies}. An application to a real dataset is shown in Section \ref{sec:An Application}. Related works to our approach are discussed in Section \ref{sec:Relatedworks}. Section \ref{sec:Discussion} concludes the paper and points out some directions worthy of further study. 

\setcounter{equation}{0} 
\section{The target-decoy approach}
\label{sec: section 2}

\subsection{Problem formulation}
\label{sec:Problem Formulation}
Consider a two-groups (case and control) study involving $m$ random variables, $X_1,X_2,\cdots,X_m$. For each random variable $X_j$ where $1 \leq j \leq m$, there are $n$ random samples $X_{j_1},X_{j_2},\cdots,X_{j_n}$, in which $X_{j_1},X_{j_2},\cdots,X_{j_{n_1}}$ are from the $n_1$ cases and $X_{j_{n_1+1}},\cdots,X_{j_n}$ are from the  $n_0 = n - n_1$ controls.

The goal is to search for random variables differently distributed between cases and controls. The null hypothesis for random variable $X_j$ used here is the exchangeable hypothesis $H_{j0}$: the joint distribution of $X_{j_1},X_{j_2},\cdots,X_{j_n}$ is symmetric. In other words, the joint probability density function of $X_{j_1},X_{j_2},\cdots,X_{j_n}$ (or the joint probability mass function if $X_{j_1},X_{j_2},\cdots,X_{j_n}$ are discrete) satisfies $f_{X_{j_1},\cdots,X_{j_n} } (x_{j_1},\cdots,x_{j_n} )=f_{X_{j_1},\cdots,X_{j_n} } \\({\pi}_n (x_{j_1},\cdots,x_{j_n}))$ for any possible $x_{j_1},\cdots,x_{j_n}$ and any permutation ${\pi}_n$ of $x_{j_1},\cdots,x_{j_n}$. If $X_{j_1},\cdots,X_{j_n}$ are independent, this hypothesis is equivalent to that $X_{j_1},\cdots,X_{j_n}$ are identically distributed. Here we use the exchangeable hypothesis to deal with the case where $X_{j_1},\cdots,X_{j_n}$ are correlated but still an exchangeable sequence of random variables \citep{chow2012probability}.

\par Let $S(x_1,x_2,\cdots,x_n)$ be some scoring function satisfying 
\begin{equation*}
S(x_1,\cdots,x_n )=S({\pi}_{n_1} (x_1,\cdots, x_{n_1} ),{\pi}_{n_0 }(x_{n_1+1},\cdots,x_n ))
\end{equation*}
for any possible $x_1,\cdots,x_n$, any permutation of $n_1$ elements ${\pi}_{n_1}(\cdot)$ and that of $n_0$ elements ${\pi}_{n_0}(\cdot)$. Note that most scoring functions evaluating the difference between $x_1,x_2,\cdots,x_{n_1}$ and $x_{n_1+1},x_{n_1+2},\cdots,x_n$ have the above symmetry property, including commonly used test statistics, e.g., the $t$-value as we used in this paper. Without loss of generality, we assume that larger scores are more significant. Note that neither the null distributions of scores nor the distributions of random variables are required to be known.

\subsection{The simplified target-decoy procedure}
\label{sec:simplified target-decoy procedure}
We first introduce the simplified version of our target-decoy procedure for FDR control.
\par \noindent \emph{Algorithm 1: the simplified target-decoy procedure}
\begin{enumerate}
\item	 For each $1 \leq j \leq m$, calculate $t$ scores including a target score and $t-1$ decoy scores. The target score is  $S_j^T=S(X_{j_1},X_{j_2},\cdots,X_{j_n})$. Each decoy score is obtained by first sampling a permutation ${\pi}_n$ of $X_{j_1},X_{j_2},\cdots,X_{j_n}$ randomly and then calculating the score as $S({\pi}_n(X_{j_1},\\X_{j_2},\cdots,X_{j_n}))$. Sort these $t$ scores in descending order. For equal scores, sort them randomly with equal probability.   
\item For each test $j$, calculate a final score $S_j$ and assign it a label $L_j\in \{T,D\}$, where $T$ and $D$ stand for target and decoy, respectively. Assume that the rank of $S_j^T$ is $i$. If $i<(t+1)/2$, let $L_j$ be $T$ and set $S_j$ as $S_j^T$. If $i>(t+1)/2$, let $L_j$ be $D$ and set $S_j$ as the score ranking $i-\lceil t/2 \rceil$. Otherwise, $i=(t+1)/2$, let $L_j$ be $T$ or $D$ randomly and set $S_j$ as $S_j^T$.
\item	 Sort the $m$ tests in descending order of the final scores. Let $i_1,\cdots,i_m$ be such that $S_{i_1} \geq \cdots \geq S_{i_m}$ (with tied values randomly broken). Let $L_{(1)},\cdots,L_{(m)}$ be the the corresponding labels $L_{i_1}, \cdots, L_{i_m}$, respectively.
\item	 If the specified FDR control level is $\alpha$, let
\begin{equation}\label{STD_Con}
K=\max\{k \big| \frac{\# \{L_{(j)}=D,j\leq k\}+1}{\# \{L_{(j)}=T,j\leq k \}\lor 1}\leq \alpha\}
\end{equation}
and reject the hypothesis with rank $j$ if $L_{(j)}=T$ and $j \leq K$.
\end{enumerate}

\begin{figure}[h!]
\includegraphics [angle=0, scale=0.48]{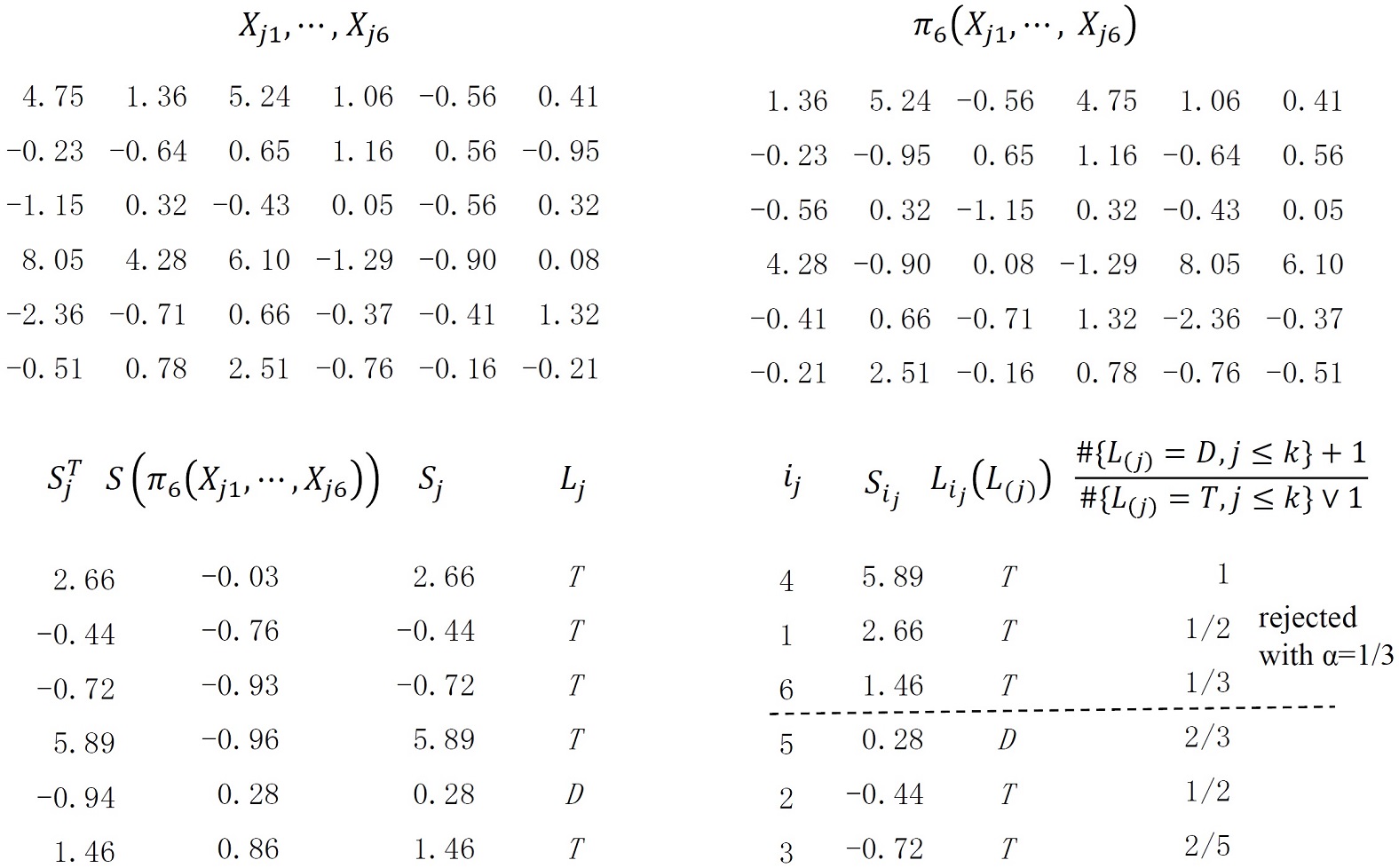}\par
\caption{An example of the simplified target-decoy procedure.}
\label{fig:TD}
\end{figure}

An example of the simplified target-decoy procedure is shown in Figure \ref{fig:TD}. In it, $m = n = 6$, and $t=2$. The first three columns of the data are from cases and the other three columns are from controls. The scoring function used is \emph{t}-value. For each row, a target score $S_j^T$ is first calculated for the original samples. Then, the procedure performs one  permutation ${\pi}_6$ and calculates one decoy score $S({\pi}_6(X_{j_1},\cdots,X_{j_6}))$, since $t-1=1$. If $S_j^T> S({\pi}_6(X_{j_1},\cdots,X_{j_6}))$,  the final score $S_j$ is set as $S_j^T$ and $L_j$ is set as $T$. Otherwise, if $S_j^T< S({\pi}_6(X_{j_1},\cdots,X_{j_6}))$, $S_j$ is set as $S({\pi}_6(X_{j_1},\cdots,X_{j_6}))$ and $L_j$ is set as $D$. The 6 tests are sorted in descending order of $S_j$ to derive $i_j$, $S_{i_j}$ and $L_{i_j}$ (i.e. $L_{(j)}$). For example, $i_1$ is $4$ because $S_4$ is maximal in all the final scores. Then, with $L_{(1)},\cdots,L_{(6)}$, we can calculate $\frac{\# \{L_{(j)}=D,j\leq k\}+1}{\# \{L_{(j)}=T,j\leq k\}\lor 1}$ for each row $k$. 
If $\alpha$ is set as $1/3$, we reject the first three hypotheses since $\frac{\# \{L_{(j)}=D,j\leq 3\}+1}{\# \{L_{(j)}=T,j\leq 3\}\lor 1} = 1/3$ and the formula is larger than $1/3$ for any $k > 3$.

\par Section \ref{sec:Theoretic Analysis} will show that the simplified target-decoy procedure controls the FDR. The random permutation used in our procedures can be generated by simple random sampling either with or without replacement, just as in the permutation tests. Similarly, with larger sampling number $t - 1$, the power of our approach will become slightly stronger as shown in Section \ref{sec:Simulation Studies}. We can set $t$ as $\min\{\binom{n}{n_0},\tau\}$, where $\tau$ is the maximum number of permutations we would perform.

Unlike other FDR control methods, our approach does not depend on the null distribution. The number of permutations, $t - 1$ can be much smaller than that used in permutation tests. In our simulations, $t-1$ was set as $49$ or $1$, while in the real data experiments, it was set as $19$. Simulations demonstrate that the target-decoy approach can still control the FDR even if $t - 1$ was set as $1$, in which case little information was revealed about the null distribution.

\subsection{The standard target-decoy procedure}
\label{sec:The target-decoy approach}
\par The +1 in the numerator of equation (\ref{STD_Con}) is essential to accomplish FDR control. However, it has a side effect of reducing the power. This effect can be amplified under some conditions, e.g., when the number of false null hypotheses or the total number of hypotheses is small. To enhance the power, we introduce a parameter $r$ into the procedure. For any fixed $1 \leq r \leq \binom{n}{n_0}$, the standard target-decoy procedure (we will omit the word \emph{standard} below for simplicity) is as follows.
\par \noindent \emph{Algorithm 2: the target-decoy procedure (Steps 1,3 are identical to Algorithm 1 and are omitted here.)}
\begin{enumerate}
\item	[(2)]	 For each $1 \leq j \leq m$, let ${\Lambda}_j=i-P_j$ where $i$ is the rank of $S_j^T$ in the $t$ scores, and $P_j$ is a random draw from uniform$[0,1)$ distribution. Calculate a final score $S_j$ and assign a label $L_j\in \{T,D,U\}$, where $T,D$ and $U$ stand for target, decoy and unused, respectively.  If ${\Lambda}_j\leq \frac{t}{2r}$, let $L_j=T$ and $S_j=S_j^T$. If $\frac{t}{2}<{\Lambda}_j \leq t$, let ${\Lambda}_j^{'}$ be a random draw from uniform$\big(0,\frac{t}{2r}\big]$ distribution, $L_j$ be $D$ and $S_j$ be the score ranking $\lceil{\Lambda}_j^{'}\rceil$-th. Otherwise, let $L_j$ be $U$ and $S_j$ be $- \infty$.

\item	[(4)] If the specified FDR control level is $\alpha$, let
\begin{equation}\label{ATD_Con}
K=\max\{k \big| \frac{1}{r}\times \frac{\# \{L_{(j)}=D,j\leq k\}+1}{\# \{L_{(j)}=T,j\leq k \}\lor 1}\leq \alpha\}
\end{equation}
and reject the hypothesis with rank $j$ if $L_{(j)}=T$ and $j \leq K$.
\end{enumerate}
\par Section \ref{sec:Theoretic Analysis} will show that the above target-decoy procedure controls the FDR for any fixed $r$. 
In practice, one can set the value of $r$ empirically or simply set $r=1$, which reduces the target-decoy procedure into its simplified version described in Section \ref{sec:simplified target-decoy procedure}. Alternatively, an algorithm can be used to choose $r$ adaptively for a given dataset as discussed in Section \ref{sec:Automatically choosing $r$}.

\subsection{The adaptive target-decoy procedure}
\label{sec:Automatically choosing $r$}
The parameter $r$ is for adjusting the probability that a true null hypothesis is labelled as $T$. On the one hand, equation (\ref{ATD_Con}) can be too conservative for a small $r$, e.g. 1 as in the simplified target-decoy procedure, because of the addition of 1 in the numerator if there are only a few false null hypotheses. For example, assume that the total number of tests is $80$ and the FDR control level is $0.01$. If $r$ is set as $1$, no hypothesis will be rejected, because the numerator of equation (\ref{ATD_Con}) is always no less than $1$ and the fraction is greater than $1/80>0.01$. On the other hand, if $r$ is too large, many false null hypotheses will be labelled as $U$ or $D$, potentially decreasing the power of testing. Thus, $r$ should be set appropriately in practice to enhance the power. Below, we provide an adaptive procedure to choose a suitable $r$ for the given dataset and the FDR control level.


\par \noindent \emph{Algorithm 3: the adaptive target-decoy procedure}
\begin{enumerate}
\item Divide the samples of each random variable into two parts as follows. Choose a suitable $n_{2}$ which is smaller than $n_0$ and $n_1$ from some range, say $5 \leq  n_2 \leq \min\{\lfloor n_0/2 \rfloor,\lfloor n_1/2 \rfloor\}$. For each random variable $X_j$ where $1 \leq j \leq m$, randomly choose $n_2$ samples from $X_{j_1},X_{j_2},\cdots,X_{j_{n_1}}$ and  $X_{j_{n_1+1}},\cdots,X_{j_n}$, respectively. Let $X^\text{1}_{j_{1}},X^\text{1}_{j_{2}},\cdots,\\X^\text{1}_{j_{2n_2}}$ be these samples. The rest has $n_1 - n_2$ samples from the cases and $n_0 - n_2$ samples from the controls. Let $X^\text{2}_{j_1},X^\text{2}_{j_2},\cdots,X^\text{2}_{j_{n - 2n_2}}$ be the rest samples.
\item Set $t$ as $\binom{2n_2}{n_2}$ and perform the target-decoy procedure on $X^\text{1}_{j_{1}},X^\text{1}_{j_{2}},\cdots,\\ X^\text{1}_{j_{2n_2}}$ where $1 \leq j \leq m$ for some range of $r$, say $R = \{1,2,5,10,15,20,\\25\}$. Let $r_{max}$ be the one such that the most hypotheses are rejected by the target-decoy procedure. 
\item Perform the target-decoy procedure on $X^\text{2}_{j_{1}},X^\text{2}_{j_{2}},\cdots,X^\text{2}_{j_{{n - 2n_2}}}$ where $1 \leq j\leq m$ with $r = r_{max}$ and reject corresponding hypotheses.
\end{enumerate}
\subsection{Control theorem}
\label{sec:Theoretic Analysis}
\par In this section, we will show that the target-decoy procedure controls the FDR. Let $H_{j} = 0$ and $H_{j} = 1$ denote that the null hypothesis for test $j$ is true and false, respectively. Note that $H_{1},H_{2},\cdots,H_{m}$ are constants in the setting of hypothesis testing. Define $Z_j$ for $1\leq j \leq m$ as follows. 
\begin{center}
\begin{tabular}{rcc}
 & $L_{j} = T$ & $L_{j} = D$ \\\hline
$H_{j} = 0$ & $Z_j=1$ &  $Z_j=-1$ \\\hline
$H_{j} = 1$ & $Z_j=0$ &  $Z_j=-2$ \\\hline
\end{tabular}
\end{center}

Let $S_{(1)},S_{(2)},\cdots,S_{(m)}$ denote the sorted scores and $Z_{(1)},Z_{(2)},\cdots,Z_{(m)}$ denote the sorted sequence of $Z_{1},Z_{2},\cdots,Z_{m}$. Let $\vv{S}$ and $\vv{S_{\not =  j}}$ denote $S_1,\cdots,S_m$ and $S_1,\cdots,S_{j-1},S_{j+1},\cdots,S_m$, respectively. Let $\vv{S_{(\cdot)}}$ and $\vv{S_{(\not =  j)}}$ denote $S_{(1)},\cdots,S_{(m)}$ and $S_{(1)},\cdots,S_{(j-1)},S_{(j+1)},\cdots, S_{(m)}$, respectively. We define $\vv{s}$, $\vv{s_{\not =  j}}$, $\vv{s_{(\cdot)}}$ and $\vv{s_{(\not =  j)}}$ similarly. For example, we will use $\vv{s_{(\cdot)}}$ to denote a sequence of $m$ constants, $s_{(1)},\cdots,s_{(m)}$, which is one of the observed values of $S_{(\cdot)}$. We also define $\vv{L},\vv{Z},\vv{H},\vv{L_{(\not =  j)}},$ etc. Then we have the following three theorems.

\begin{theorem}\label{thm:equalprobability}
In the simplified target-decoy procedure, if the $m$ random variables are independent, then for any fixed $1 \leq j \leq m$ and any possible $\vv{s_{(\cdot)}}$ and $\vv{z_{(\not = j)}}$ we have
\begin{equation*}\begin{aligned}
&\Pr\Big(Z_{(j)} = -1 \big |\vv{S_{(\cdot)}} =\vv{s_{(\cdot)} },\vv{Z_{(\not = j)}}=\vv{z_{(\not = j)}}\Big) \\= &\Pr\Big(Z_{(j)} = 1 \big |\vv{S_{(\cdot)}} =\vv{s_{(\cdot)} },\vv{Z_{(\not = j)}}=\vv{z_{(\not = j)}}\Big).
\end{aligned}\end{equation*}
\end{theorem}

\begin{theorem}\label{thm:ratio}
In the target-decoy procedure, if the $m$ random variables are independent, then for any fixed $1 \leq j \leq m$ and any possible $\vv{s_{(\cdot)}}$ and $\vv{z_{(\not = j)}}$ we have
\begin{equation*}\begin{aligned}
&\Pr\Big(Z_{(j)} = -1 \big |\vv{S_{(\cdot)}} =\vv{s_{(\cdot)} },\vv{Z_{(\not = j)}}=\vv{z_{(\not = j)}}\Big) \\= & r\Pr\Big(Z_{(j)} = 1 \big |\vv{S_{(\cdot)}} =\vv{s_{(\cdot)} },\vv{Z_{(\not = j)}}=\vv{z_{(\not = j)}}\Big).
\end{aligned}\end{equation*}
\end{theorem}

\begin{theorem}\label{thm:fdrcontrol}
Suppose that $S_{(1)},S_{(2)},\cdots,S_{(m)}$,$Z_{(1)},Z_{(2)},\cdots,Z_{(m)}$ are random variables satisfying $S_{(1)}\geq S_{(2)}\geq\cdots\geq S_{(m)}$ and $Z_{(1)},Z_{(2)},\cdots,Z_{(m)} \in \{-2,-1,0,1\}$, and $r$ is a positive constant. For any $\alpha \in (0,1]$, define
\begin{equation*}\label{TheoremC_10}
K=\max\{k \big| \frac{1}{r}\times\frac{\# \{Z_{(j)}<0 , j\leq k\}+1}{\# \{Z_{(j)}\geq 0,j\leq k\}\lor 1}\leq \alpha\}.
\end{equation*}
If there is no such $k$, let $K=0$. If for any fixed $j$ and any possible $\vv{s_{(\cdot)}}$ and $\vv{z_{(\not = j)}}$,
\begin{equation}\begin{aligned}\label{TheoremC_11}
&\Pr\Big(Z_{(j)} = -1 \big |\vv{S_{(\cdot)}} =\vv{s_{(\cdot)} },\vv{Z_{(\not = j)}}=\vv{z_{(\not = j)}}\Big)
\\= & r\Pr\Big(Z_{(j)} = 1 \big |\vv{S_{(\cdot)}} =\vv{s_{(\cdot)} },\vv{Z_{(\not = j)}}=\vv{z_{(\not = j)}}\Big),
\end{aligned}\end{equation}
then we have
\begin{equation*}
\mathbb{E}\bigg(\frac{\#\{Z_{(j)}=1, j\leq K\}}{\#\{Z_{(j)}\geq 0, j\leq K\}\lor 1}\bigg)< \alpha.
\end{equation*}
\end{theorem}

The proofs of these theorems are given in the Supplementary Materials. Theorem \ref{thm:fdrcontrol} indicates that the target-decoy procedure controls the FDR if the $m$ random variables are independent. 

Specially, all of the above theorems hold for the adaptive target-decoy procedure. Recall that the null hypothesis for random variable $X_j$ used here is the exchangeable hypothesis $H_{j0}$:  the joint probability density function of $X_{j_1},X_{j_2},\cdots,X_{j_n}$ satisfies $f_{X_{j_1},\cdots,X_{j_n} } (x_{j_1},\cdots,x_{j_n} )=f_{X_{j_1},\cdots,X_{j_n} } ({\pi}_n (x_{j_1},\cdots,\\x_{j_n}))$ for any possible $x_{j_1},\cdots,x_{j_n}$ and any permutation ${\pi}_n$ of $x_{j_1},\cdots,x_{j_n}$. If  $H_{j0}$ is true, it is easy to see that $X^\text{2}_{j_1},X^\text{2}_{j_2},\cdots,X^\text{2}_{j_{n - 2n_2}}$ are also exchangeable.

\setcounter{equation}{0} 
\section{Simulation Studies}
\label{sec:Simulation Studies}
We used Monte-Carlo simulations to study the performance of our approach. The target-decoy procedure were compared with two popular traditional multiple testing methods, including the Bayes method \citep{storey2002direct, storey2003positive, storey2004strong} and the empirical Bayes method \citep{efron2001empirical, efron2002empirical, efron2008microarrays}. Simulations were conducted for both independent and dependent random variables. We mainly evaluated the performance of the simplified target-decoy procedure. To show the effectiveness of adjusting $r$, we also did a simulation on a small dataset and compared the adaptive target-decoy procedure with the simplified target-decoy procedure. 

\subsection{Simulation setup}
In the simulation, we considered the case-control studies in which the random variables follow the normal distribution or the gamma distribution. In addition to the normal distribution, we did simulation experiments for the gamma distribution because many random variables in real world are gamma-distributed. Recall that the case-control study consists of $m$ random variables. For each random variable, there are $n$ random samples, $n_1$ of which are from the cases and the other $n_0 = n - n_1$ are from the controls. Let $X_{j_1},X_{j_2},\cdots,X_{j_n}$ be the $n$ random samples for random variable $X_j$.

\par The observation values from the normal distribution were generated in a way similar to \citet*{benjamini2006adaptive}. First, let ${\zeta}_0,{\zeta}_{11}, \cdots,{\zeta}_{1n},\cdots,{\zeta}_{m1}, \cdots,{\zeta}_{mn}$ be independent and identically distributed random variables following the $N(0,1)$ distribution. Next, let $X_{ji} = \sqrt{\rho}{\zeta}_0 + \sqrt{\rho}{\zeta}_{ji} + {\mu}_{ji}$ for $j = 1,\cdots,m$ and $i = 1,\cdots,n$. We used $\rho=0, 0.4$ and $0.8$, with $\rho=0$ corresponding to independence and $\rho=0.4$ and $0.8$ corresponding to typical moderate and high correlation values estimated from real microarray data, respectively \citep{almudevar2006utility}. The values of ${\mu}_{ji}$ are zero for $i = n_1+1,n_1+2,\cdots,n$, the $n_0$ controls. For the $n_1$ cases where $i = 1,2,\cdots,n_1$, the values of ${\mu}_{ji}$ are also zero for $j = 1,2,\cdots,m_0$, the $m_0$ hypotheses that are true null. The values of ${\mu}_{ji}$ for $i = 1,2,\cdots,n_1$ and $j=m_0 + 1,\cdots,m$ are set as follows. We let ${\mu}_{ji} =1,2, 3$ and $4$ for $j=m_0 + 1,m_0 + 2,m_0 +3,m_0 +4$, respectively. Similarly, we let ${\mu}_{ji} =1,2, 3$ and $4$ for $j=m_0 + 5,m_0 + 6,m_0 +7,m_0 +8$, respectively. This cycle was repeated to produce ${\mu}_{(m_0 + 1)1},\cdots,{\mu}_{(m_0 + 1)n_1},\cdots,{\mu}_{m1},\cdots,{\mu}_{mn_1}$ for the false null hypotheses.
\par The observation values from the gamma distribution, which is characterized using shape and scale, were generated in the following way. First, let ${\Gamma}_0,{\Gamma}_{11}, \cdots,{\Gamma}_{1n},\cdots,{\Gamma}_{m1}, \cdots,{\Gamma}_{mn}$ be independent random variables where $\Gamma_0$ follows the ${\Gamma}(k_0,1)$ distribution and $\Gamma_{ji}$ follows the ${\Gamma}(k_{ji},1)$ distribution for any $j = 1,\cdots,m$ and $i = 1,\cdots,n$. Next, let $X_{ji} = \Gamma_{ji}$ for $j = 1,\cdots,m$ and $i = 1,\cdots,n$ in the simulation study for independent random variables and let $X_{ji} = \Gamma_0 + \Gamma_{ji}$ for dependent random variables. To obtain reasonable correlation values, $k_0$ was set as 4 and $k_{ji}$ was set as 1 for $i = n_1+1,n_1+2,\cdots,n$, the $n_0$ controls. For the $n_1$ cases where $i = 1,2,\cdots,n_1$, $k_{ji}$ was set as 1 for $j = 1,\cdots,m_0$, the $m_0$ hypotheses that are true null. The values of $k_{ji}$ for $i = 1,2,\cdots,n_1$ and $j=m_0 + 1,\cdots,m$ are set as follows. We let $k_{ji} =2,3, 4$ and $5$ for $j=m_0 +1,m_0 +2,m_0 +3,m_0 +4$, respectively. Similarly, we let $k_{ji} =2,3, 4$ and $5$ for $j=m_0 + 5,m_0 + 6,m_0 +7,m_0 +8$, respectively. This cycle was repeated to produce $k_{(m_0 + 1)1},\cdots,k_{(m_0 + 1)n_1},\cdots,k_{m1},\cdots,k_{mn_1}$ for the false null hypotheses.

\par The specified FDR control level $\alpha$ was set as 5$\%$ or 10$\%$. The total number of tests, $m$, was set as 10000. The proportion of false null hypotheses was $1\%$ or $10\%$. The total sample size, $n$, was set as 20, consisting of the same numbers of cases and controls. 


\par Three different approaches to FDRs were compared, including the Bayes method \citep{storey2002direct, storey2003positive, storey2004strong}, the empirical Bayes method \citep{efron2001empirical, efron2002empirical, efron2008microarrays} and our target-decoy approach. The Bayes method and the empirical Bayes method are among the most remarkable multiple testing methods. To compare the power of these methods, we rejected the hypotheses against the specified FDR control level $\alpha$. The rejection threshold, $s$, for the Bayes method was set as the largest $p$-value such that $q$-value($s$) is no more than $\alpha$ \citep{storey2002direct, storey2003positive}. The rejection threshold, $s$, for the empirical Bayes method was set as the minimum $z$-value such that Efdr($s$) is no more than $\alpha$, where Efdr($s$) is the expected fdr of hypotheses with $z$-values no smaller than $s$ \citep{efron2007size, efron2004large}. Specifically, the R packages "locfdr" version 1.1-8 \citep{efron2004large}, and "qvalue" version 2.4.2 \citep{storey2003statistical} were used. Each simulation experiment was repeated for 1000 times. We calculated the mean number of rejected hypotheses to evaluate the power of each method. The FDRs of rejected hypotheses were calculated by the means of false discovery proportions (FDPs). Note that the variance of the mean of FDPs of 1000 repetitions is one thousandth of the variance of FDPs. We also estimated the standard deviation of the mean of FDPs from the sample standard deviation of FDPs.

\par The $p$-values of the Bayes method and the $z$-values of the empirical Bayes method were calculated with the Student's $t$-test, Wilcoxon rank sum test or the Student's $t$-test with permutation. For the permutation method, we sampled the cases and the controls for each test, calculated the $z$-values for sampled data by $t$-test, and calculated the $p$-values with the null distribution of pooled $z$-values \citep{xie2005note,liu2014phase}. The sampling number of permutations was set as 10 \citep{efron2012large}.

For our target-decoy approach, the cases and the controls of each test were permuted for 49 times or only once, and the $t$-values and the test statistics of the Wilcoxon rank sum test were used. We did the one-permutation experiments where little information about the null distributions was revealed to demonstrate that our approach does not rely on the null distribution. Because the permutation is performed inherently in our target-decoy approach, the extra permutation is unnecessary. 


\par We will use abbreviations to represent the experiments. For example, \emph{Bayes,permutation,Normal,10\%,$\rho = 0.8$} represents the simulation experiment where the Bayes method combined with the pooled permutation is used, the random variables follow the normal distribution, the proportion of false null hypotheses is $10\%$ and the correlation values are 0.8. For our target-decoy approach, \emph{$t$-value,49,Gamma,1\%} represents the simulation experiment where the $t$-value is used as the score, 49 permutations are performed for each test, the random variables follow the gamma distribution and the proportion of false null hypotheses is as low as $1\%$.

\subsection{Results on independent random variables}
\begin{figure}[h!]
\includegraphics [angle=0, scale=0.33]{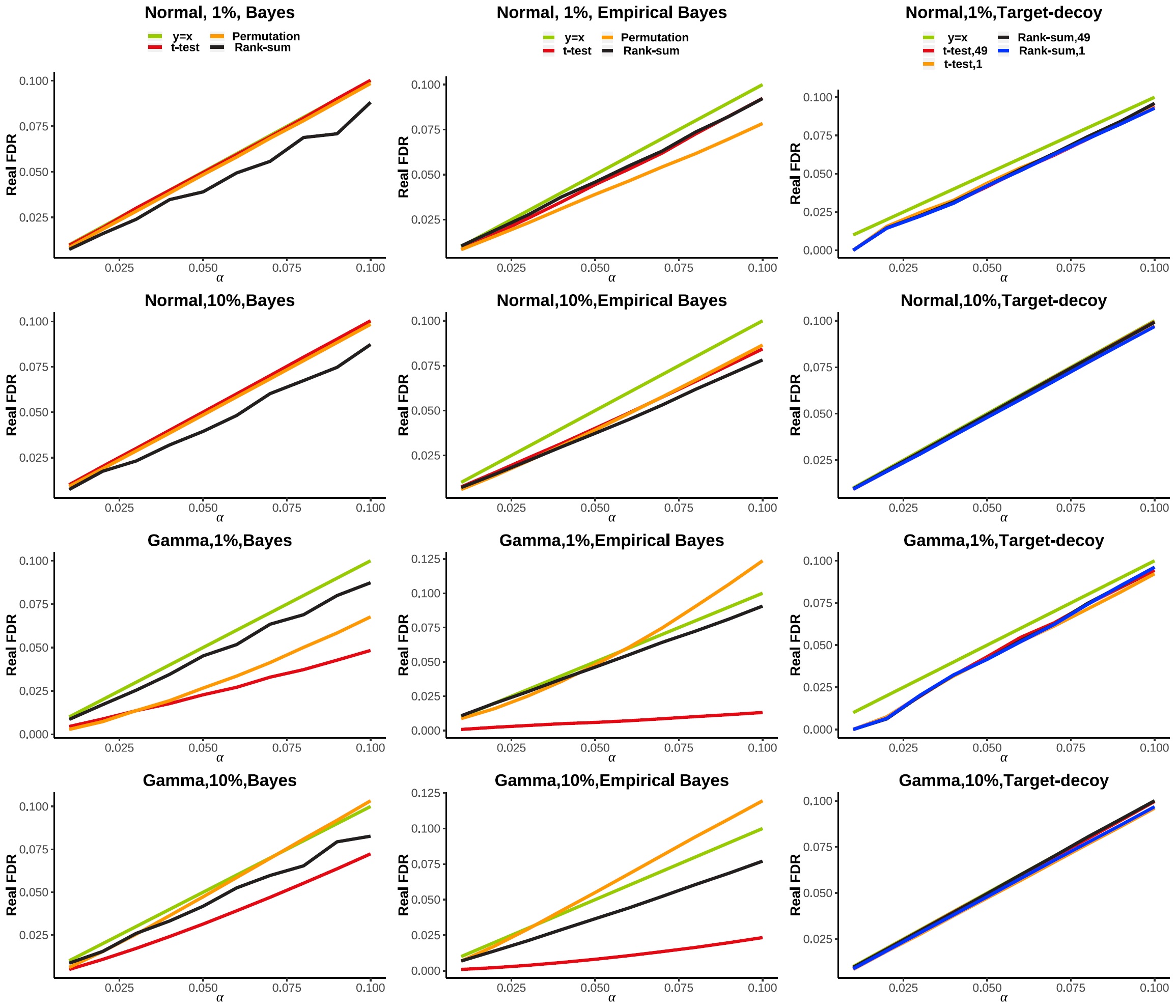}\par
\caption{Real FDRs with independent random variables.}
\label{fig:FDR_ind}
\end{figure}

Figure~\ref{fig:FDR_ind} shows the real FDRs of different methods with independent random variables while the specified FDR control level $\alpha$ was no more than $10\%$. Table~\ref{tab:s20_1} gives the real FDRs  while the specified FDR control level $\alpha$ was $5\%$ or $10\%$. In all cases, the target-decoy approach controlled the FDR, and the real FDRs were favourably close to $\alpha$. The empirical Bayes and Bayes methods performed well when the random variables followed the normal distribution. However, they considerably overestimated the FDRs with $t$-test for the gamma distribution. With the pooled permutation, some of the real FDRs exceeded $\alpha$ for the gamma distribution as marked by asterisks in the table. Of course, some small exceedances are not necessarily the evidence of a fail of FDR control but may be due to Monte Carlo error. At last, the Wilcoxon rank-sum test coupled with Bayes or empirical Bayes occasionally overestimated the FDRs.

\begin{table}[!ht]
\caption{Real FDRs with independent random variables. The FDRs were calculated as the means of FDPs and the standard deviations of the means are less than $0.0020$. All the cases where FDRs exceed the control level $\alpha$ are labelled with $\ast$. \label{tab:s20_1}}
\begin{center}
\renewcommand\arraystretch{0.8}
\scalebox{0.95}{
\begin{tabular}{|r | cc| cc|cc| cc|}\hline
		& \multicolumn{2}{c|}{Normal,1\%} & \multicolumn{2}{c|}{Normal,10\%}& \multicolumn{2}{c|}{Gamma,1\%}& \multicolumn{2}{c|}{Gamma,10\%}\\
		\hline
$\alpha$ &	0.05 & 0.10 & 0.05 & 0.10 & 0.05 & 0.10 & 0.05 & 0.10 \\\hline
\multicolumn{9}{|c|}{ Bayes}\\\hline
$t$-test  & 0.050  & 0.100  & 0.050  & 0.100  & 0.023  & 0.048  & 0.031  & 0.072   \\
permutation  & 0.048  & 0.099  & 0.048  & 0.098  & 0.027  & 0.068  & 0.047  & 0.103$^\ast$  \\
rank-sum & 0.039  & 0.088  & 0.039  & 0.087  & 0.045  & 0.087  & 0.042  & 0.083  \\\hline
\multicolumn{9}{|c|}{Empirical Bayes}\\\hline
$t$-test  & 0.044  & 0.092  & 0.040  & 0.084  & 0.006  & 0.013  & 0.008  & 0.023  \\
permutation & 0.039  & 0.078  & 0.039  & 0.086  & 0.048  & 0.124$^\ast$  & 0.055$^\ast$  & 0.119$^\ast$ \\
rank-sum & 0.046  & 0.092  & 0.037  & 0.078  & 0.046  & 0.091  & 0.037  & 0.077  \\\hline
\multicolumn{9}{|c|}{Target-decoy}\\\hline
$t$-value,49  & 0.041  & 0.094  & 0.049  & 0.099  & 0.043  & 0.094  & 0.050  & 0.100   \\
$t$-value,1  & 0.044  & 0.093  & 0.048  & 0.097  & 0.042  & 0.092  & 0.047  & 0.096   \\
rank-sum,49  & 0.042  & 0.096  & 0.049  & 0.099  & 0.042  & 0.096  & 0.050  & 0.100   \\
rank-sum,1  & 0.042  & 0.093  & 0.048  & 0.097  & 0.042  & 0.096  & 0.048  & 0.097  \\\hline
\end{tabular}
}
\end{center}
\end{table}

\begin{table}[!ht]
\caption{Power with independent random variables. All the cases where FDRs exceed $\alpha$ as shown in Table \ref{tab:s20_1} are labelled with $\ast$.
\label{tab:s20_2}}
\begin{center}
\renewcommand\arraystretch{0.8}
\scalebox{1}{
\begin{tabular}{|r | cc | cc | cc | cc|}\hline
		& \multicolumn{2}{c|}{Normal,1\%} & \multicolumn{2}{c|}{Normal,10\%}& \multicolumn{2}{c|}{Gamma,1\%}& \multicolumn{2}{c|}{Gamma,10\%}\\
		\hline
$\alpha$ &	0.05 & 0.10 & 0.05 & 0.10 & 0.05 & 0.10 & 0.05 & 0.10 \\\hline
\multicolumn{9}{|c|}{Bayes}\\\hline
$t$-test    & 71    & 80    & 845   & 937   & 40    & 50    & 687   & 798  \\
permutation    & 71    & 80    & 842   & 933   & 41    & 55    & 737   & 861$^\ast$  \\
rank-sum   & 67    & 76    & 813   & 906   & 48    & 59    & 734   & 836   \\\hline
\multicolumn{9}{|c|}{Empirical Bayes}\\\hline
$t$-test    & 70    & 78    & 823   & 909   & 23    & 32    & 534   & 650  \\
permutation   & 69    & 76    & 821   & 913   & 49    & 66$^\ast$    & 755$^\ast$   & 891$^\ast$   \\
rank-sum     & 69    & 77    & 806   & 889   & 47    & 59    & 715   & 823  \\\hline
\multicolumn{9}{|c|}{Target-decoy}\\\hline
$t$-value,49   & 69    & 79    & 843   & 935   & 45    & 60    & 743   & 853  \\
$t$-value,1   & 69    & 79    & 841   & 931   & 45    & 60    & 736   & 845  \\
rank-sum,49  & 67    & 77    & 834   & 926   & 42    & 60    & 755   & 872  \\
rank-sum,1 & 66    & 77    & 831   & 922   & 42    & 60    & 751   & 865  \\\hline
\end{tabular}
}
\end{center}
\end{table}

Table~\ref{tab:s20_2} shows the statistical powers of different methods with independent random variables. When the random variables followed the normal distribution, the powers of the three methods were overall comparable with each other. In the case of gamma distribution, the target-decoy approach was much more powerful than Bayes and empirical Bayes when $t$-test was used. Permutation based Bayes and empirical Bayes had higher power but at the cost of uncontrolled FDR. When the Wilcoxon rank-sum test was used, our approach was more powerful than the other two methods except the only case of \emph{Gamma,1\%} and $alpha$=0.05. 


\par In all the above experiments, the target-decoy approach successfully controlled the FDR and meanwhile it was remarkably powerful. Notably, the results obtained with 49 permutations or 1 permutation in the target-decoy approach were quite similar, indicating that the proposed approach is not sensitive to the number of permutations. 

\subsection{Results on dependent random variables}
\label{sec:dependent}
In this part, we present the simulation results for the simplified target-decoy procedure on dependent random variables. Table~\ref{tab:s20_3} shows the real FDRs of different methods with dependent random variables while the specified FDR control level $\alpha$ was $5\%$ or $10\%$. The results show that the $t$-test with empirical Bayes overestimated the FDRs for the gamma distribution. The real FDRs of pooled permutation significantly exceeded $\alpha$ when the random variables followed the gamma distribution. The Wilcoxon rank-sum test with Bayes or empirical Bayes overestimated the FDRs. The target-decoy approach controlled the FDR in all cases.

Table~\ref{tab:s20_4} shows the statistical power of different methods with dependent random variables. When the random variables followed the normal distribution, the Bayes method was less powerful than the target-decoy approach while the Wilcoxon rank-sum test was used. Though the Bayes method seems to be a little more powerful than the target-decoy approach while the $t$-test was used, the real FDR of this method exceeded the specified FDR control level. The empirical Bayes method was less powerful than the Bayes method and our target-decoy approach in the \emph{Normal,10\%,$\rho = 0.4$} experiments.

\par When the random variables followed the gamma distribution, the target-decoy approach was much more powerful than the Bayes and empirical Bayes methods, even if only one permutation was performed. Though the pooled permutation seems to be powerful, the FDRs were not controlled.

Similar to the results for independent random variables, the target-decoy approach performed significantly better than other methods for dependent random variables. It controlled the FDR in all cases without loss of statistical power.

\begin{table}
\caption{Real FDRs with dependent random variables. The FDRs were calculated as the means of FDPs and the standard deviations of the means of FDPs are less than $0.0021$. All the cases where FDRs exceed $\alpha$ are labelled with $\ast$.\label{tab:s20_3}}
\centering
\begin{center}
\footnotesize
\scalebox{0.9}{
\begin{tabular}{|>{\raggedleft}p{1.9cm}| >{\raggedleft}p{0.6cm}>{\centering}p{0.6cm}  >{\centering}p{0.6cm}>{\centering}p{0.7cm}  | >{\centering}p{0.6cm}>{\centering}p{0.6cm}  p{0.6cm}p{0.7cm} |p{0.6cm}p{0.6cm} p{0.6cm} p{0.7cm}|}\hline
    & \multicolumn{4}{c|}{Normal,$\rho =0.4$} & \multicolumn{4}{c|}{Normal,$\rho =0.8$}&  \multicolumn{4}{c|}{Gamma}\\\hline
		& \multicolumn{2}{c}{$1\%$} & \multicolumn{2}{c|}{$10\%$}& \multicolumn{2}{c}{$1\%$}& \multicolumn{2}{c|}{$10\%$}& \multicolumn{2}{c}{$1\%$} & \multicolumn{2}{c|}{$10\%$}\\
		\hline
$\alpha$ &	0.05 & 0.10 &0.05 & 0.10 &0.05 & 0.10 & 0.05 & 0.10 & 0.05 & 0.10 &0.05 & 0.10 \\\hline
\multicolumn{13}{|c|}{Bayes}\\\hline
$t$-test & 0.052$^\ast$  & 0.102$^\ast$  & 0.050  & 0.100  & 0.050  & 0.101$^\ast$  & 0.050  & 0.100  & 0.023  & 0.048  & 0.031  & 0.072  \\
permutation & 0.050  & 0.100  & 0.048  & 0.098  & 0.049  & 0.099  & 0.047  & 0.098  & 0.026  & 0.067  & 0.047  & 0.103$^\ast$ \\
rank-sum  & 0.046  & 0.088  & 0.044  & 0.085  & 0.038  & 0.092  & 0.039  & 0.083  & 0.043  & 0.085  & 0.042  & 0.082 \\\hline
\multicolumn{13}{|c|}{Empirical Bayes}\\\hline
$t$-test  & 0.047  & 0.097  & 0.044  & 0.090  & 0.048  & 0.100  & 0.047  & 0.097  & 0.006  & 0.013  & 0.008  & 0.023   \\
permutation  & 0.042  & 0.083  & 0.043  & 0.093  & 0.041  & 0.084  & 0.045  & 0.099  & 0.048  & 0.123$^\ast$  & 0.055$^\ast$  & 0.121$^\ast$   \\
rank-sum & 0.049  & 0.095  & 0.042  & 0.086  & 0.048  & 0.094  & 0.046  & 0.095  & 0.045  & 0.090  & 0.037  & 0.077 \\\hline
\multicolumn{13}{|c|}{Target-decoy}\\\hline
$t$-value,49  & 0.047  & 0.097  & 0.050  & 0.100  & 0.047  & 0.095  & 0.049  & 0.100  & 0.043  & 0.094  & 0.048  & 0.099 \\
$t$-value,1 & 0.046  & 0.096  & 0.048  & 0.098  & 0.045  & 0.096  & 0.049  & 0.100  & 0.042  & 0.092  & 0.047  & 0.096 \\
rank-sum,49 & 0.049  & 0.099  & 0.049  & 0.099  & 0.045  & 0.096  & 0.050  & 0.100  & 0.042  & 0.090  & 0.050  & 0.100\\
rank-sum,1 & 0.048  & 0.100  & 0.049  & 0.099  & 0.048  & 0.097  & 0.050  & 0.100  & 0.040  & 0.089  & 0.047  & 0.096\\\hline
\end{tabular}}
\end{center}
\end{table}

\begin{table}
\caption{Power with dependent random variables. The sample size is 20. All the cases where FDRs exceed $\alpha$ as shown in Table \ref{tab:s20_3} are labelled with $\ast$. \label{tab:s20_4}}
\begin{center}
\footnotesize
\scalebox{0.9}{
\begin{tabular}{|>{\raggedleft}p{1.9cm}| >{\raggedleft}p{0.6cm}>{\centering}p{0.6cm}  >{\centering}p{0.6cm}>{\centering}p{0.6cm}  | >{\centering}p{0.6cm}>{\centering}p{0.6cm}  p{0.6cm}p{0.6cm} |p{0.6cm}p{0.6cm} p{0.6cm} p{0.6cm}|}\hline
    & \multicolumn{4}{c|}{Normal,$\rho =0.4$} & \multicolumn{4}{c|}{Normal,$\rho =0.8$}&  \multicolumn{4}{c|}{Gamma}\\\hline
		& \multicolumn{2}{c}{$1\%$} & \multicolumn{2}{c|}{$10\%$}& \multicolumn{2}{c}{$1\%$}& \multicolumn{2}{c|}{$10\%$}& \multicolumn{2}{c}{$1\%$} & \multicolumn{2}{c|}{$10\%$}\\
		\hline
$\alpha$ &	0.05 & 0.10 &0.05 & 0.10 &0.05 & 0.10 & 0.05 & 0.10 & 0.05 & 0.10 &0.05 & 0.10 \\\hline
\multicolumn{13}{|c|}{Bayes}\\\hline
$t$-test  & 82$^\ast$    & 90$^\ast$    & 927   & 1016  & 101   & 108$^\ast$   & 1047  & 1109  & 40    & 50    & 687   & 797\\
permutation  & 82    & 90    & 922   & 1012  & 101   & 108   & 1043  & 1106  & 42    & 55    & 737   & 861$^\ast$  \\
rank-sum  & 80    & 87    & 907   & 983   & 98    & 106   & 1031  & 1086  & 47    & 59    & 735   & 836    \\\hline
\multicolumn{13}{|c|}{Empirical Bayes}\\\hline
$t$-test & 81    & 90    & 914   & 999   & 100   & 108   & 1043  & 1105  & 23    & 32    & 536   & 652  \\
permutation & 81    & 87    & 912   & 1003  & 99    & 106   & 1041  & 1108  & 49    & 66$^\ast$    & 757$^\ast$   & 893$^\ast$ \\
rank-sum & 80    & 88    & 900   & 984   & 99    & 107   & 1040  & 1102  & 47    & 59    & 716   & 823  \\\hline
\multicolumn{13}{|c|}{Target-decoy}\\\hline
$t$-value,49 & 81    & 90    & 926   & 1015  & 100   & 108   & 1046  & 1109  & 44    & 60    & 741   & 852  \\
$t$-value,1 & 81    & 89    & 923   & 1013  & 100   & 108   & 1046  & 1109  & 45    & 60    & 735   & 845  \\
rank-sum,49 & 80    & 89    & 917   & 1007  & 99    & 107   & 1045  & 1108  & 42    & 59    & 756   & 870 \\
rank-sum,1 & 80    & 89    & 916   & 1005  & 99    & 107   & 1044  & 1108  & 41    & 59    & 749   & 863  \\\hline
\end{tabular}}
\end{center}
\end{table}

\subsection{Simulation for the adaptive procedure} 
\label{sec: Simulation for adjusting $r$}
To show the effectiveness of the adaptive target-decoy procedure for small datasets, a case-control study involving $200$ random variables was simulated. The null hypotheses of 20 random variables were true and the others were false. For each random variable, there were $20$ random samples, $10$ of which were from the cases and the other $10$ were from the controls. The observation values from the cases where the null hypotheses were false followed the $N(4,1)$ distribution, and all the other observation values followed the $N(0,1)$ distribution. All the observation values were independent. In the simulation, the cases and the controls of each test were permuted for 49 times and the $t$-values were used.

As shown in Table \ref{tab:adjust $r$}, the adaptive procedure controlled the FDR for all values of $\alpha$, and its power was much larger than the simplified target-decoy procedure for small $\alpha$.

\begin{table}
\caption{Real FDRs and power of the adaptive target-decoy procedure. The FDRs were calculated by the means of FDPs of 1000 repetitions. \label{tab:adjust $r$}}
\centering
\renewcommand\arraystretch{0.7}
\begin{tabular}{|p{0.9cm}<{\raggedleft}|>{\centering}p{0.8cm} >{\centering}p{0.8cm} >{\centering}p{0.8cm} >{\centering}p{0.8cm} >{\centering}p{0.8cm} >{\centering}p{0.8cm} >{\centering}p{0.8cm} >{\centering}p{0.8cm} >{\centering}p{0.8cm} p{0.8cm}<{\centering}|}%
\hline
$\alpha$ & $0.01$ & $0.02$ & $0.03$ & $0.04$ & $0.05$ & $0.06$ & $0.07$ & $0.08$ & $0.09$ & $0.10$\\\hline
\multicolumn{11}{|c|}{Simplified target-decoy procedure}\\\hline
FDR & $0$ &  $0$ &  $0$ &  $0.006$ &  $0.044$ &  $0.044$ &  $0.044$ &  $0.055$ &  $0.070$ &  $0.087$ \\
Power & $0$ &  $0$ &  $0$ &  $1$ &  $21$ &  $21$ &  $21$ &  $21$ &  $22$ &  $22$ \\\hline
\multicolumn{11}{|c|}{Adaptive target-decoy procedure}\\\hline
FDR & $0.007$ &  $0.018$ &  $0.026$ &  $0.032$ &  $0.044$ &  $0.049$ &  $0.058$ &  $0.069$ &  $0.079$ &  $0.093$ \\
Power & $13$ &  $18$ &  $18$ &  $19$ &  $18$ &  $20$ &  $21$ &  $21$ &  $21$ &  $22$ \\\hline
\end{tabular}
\end{table}

\section{An Application to Real data}
\label{sec:An Application}
We applied the target-decoy approach to an \emph{Arabidopsis} microarray dataset. To determine whether \emph{Arabidopsis} genes respond to oncogenes encoded by the transfer-DNA (T-DNA) or to bacterial effector proteins codelivered by \emph{Agrobacteria} into the plant cells, \citet{lee2009agrobacterium} conducted microarray experiments at $3$ h and $6$ d after inoculating wounded young \emph{Arabidopsis} plants with two different \emph{Agrobacterium} strains, C58 and GV3101. Strain GV3101 is a cognate of strain C58, which only lacks T-DNA, but possesses proteinaceous virulence (Vir) factors such as VirD2, VirE2, VirE3 and VirF \citep{vergunst2003recognition}. Wounded, but uninfected, stalks were served as control. Here we just use the 6-d postinoculation data as an example (downloaded from http://www.ncbi.nlm.nih.gov/geo/, GEO accession: GSE14106). The data consisting of 22810 genes were obtained from the C58 infected and control stalks. Both infected and control stalks were with three replicates.
\par Similar to the simulation experiments, the Bayes method, the empirical Bayes method and our target-decoy approach (the simplified procedure) are compared here. The $p$-values in the Bayes method and the $z$-values in the empirical Bayes method were calculated with the Student's $t$-test, Wilcoxon rank sum test, and the Student's $t$-test with permutation, respectively. For the Bayes method, two-tailed tests were used. For the empirical Bayes method, we first transformed the FDR control level to the threshold of local fdr and then identified differentially expressed genes according to the threshold. For the target-decoy approach, the absolute $t$-values and the test statistics of the Wilcoxon rank sum test were used.

Because it is unknown which genes were really differentially expressed, the real FDRs cannot be computed here. The power of these methods are compared. In fairness, the sampling numbers were set as $19 = \binom{6}{3} - 1$ in all the experiments, including the pooled permutation and the target-decoy approach. That is, all possible permutations were generated for each gene.

As shown in Table $\ref{Table_real_data}$, no differentially expressed genes were found by the empirical Bayes method or the Wilcoxon rank-sum test. For the Bayes method, the $t$-test was more powerful than the pooled permutation for small $\alpha$ ($\leq 0.05$) while the pooled permutation was more powerful for large $\alpha$ ($\geq 0.06$). The target-decoy approach with $t$-test was most powerful for $0.04 \leq \alpha \leq 0.09$. The additional genes identified by the target-decoy approach are reliable, because similar numbers of genes, i.e., 785 genes for FDR 0.034, 1427 genes for FDR 0.050 and 2071 genes for FDR 0.065, were reported by a more specific analysis \citep{tan2014general}.

\begin{table}
\begin{threeparttable}
\renewcommand\arraystretch{0.7}
\caption{Power of different methods for \emph{Arabidopsis} microarray data.  \label{Table_real_data}}
\begin{center}
\begin{tabular}{|p{2.5cm}<{\raggedleft}|>{\centering}p{0.69cm} >{\centering}p{0.69cm} >{\centering}p{0.69cm} >{\centering}p{0.69cm} >{\centering}p{0.69cm} >{\centering}p{0.69cm} >{\centering}p{0.69cm} >{\centering}p{0.69cm} >{\centering}p{0.69cm} p{0.69cm}<{\centering}|}%
\hline
$\alpha$ & $0.01$ & $0.02$ & $0.03$ & $0.04$ & $0.05$ & $0.06$ & $0.07$ & $0.08$ & $0.09$ & $0.10$\\\hline
\multicolumn{11}{|c|}{Bayes}\\\hline
$t$-test & $0$ &  $5$ &  $5$ &  $171$ &  $322$ &  $712$ &  $1108$ &  $1469$ &  $1875$ &  $2208$ \\
permutation & $0$ &  $0$ &  $0$ &  $0$ &  $251$ &  $1266$ &  $2035$ &  $2816$ &  $3499$ &  $4150$ \\
rank-sum test & $0$ &  $0$ &  $0$ &  $0$ &  $0$ &  $0$ &  $0$ &  $0$ &  $0$ &  $0$ \\\hline
\multicolumn{11}{|c|}{Empirical Bayes}\\\hline
\emph{t}-test & $0$ &  $0$ &  $0$ &  $0$ &  $0$ &  $0$ &  $0$ &  $0$ &  $0$ &  $0$ \\
permutation & $0$ &  $0$ &  $0$ &  $0$ &  $0$ &  $0$ &  $0$ &  $0$ &  $0$ &  $0$ \\
rank-sum test & $\ast$ &  $\ast$ &  $\ast$ &  $\ast$ &  $\ast$ &  $\ast$ &  $\ast$ &  $\ast$ &  $\ast$ &  $\ast$ \\\hline
\multicolumn{11}{|c|}{Target-decoy}\\\hline
\emph{t}-value & $0$ &  $0$ &  $0$ &  $1026$ &  $1481$ &  $1824$ &  $2204$ &  $2951$ &  $3506$ &  $3820$ \\
rank-sum test & $0$ &  $0$ &  $0$ &  $0$ &  $0$ &  $0$ &  $0$ &  $0$ &  $0$ &  $0$ \\\hline
\end{tabular}
\begin{tablenotes}
\item[$\ast$] The R package `locfdr' crashed while the Wilcoxon rank-sum test is used.
\end{tablenotes}
\end{center}
\end{threeparttable}
\end{table}

\section{Related works}
\label{sec:Relatedworks}
Our approach was inspired by the widely used target-decoy database search approach to estimating the FDR of peptide identifications in tandem mass spectrometry-based proteomics ~\citep{Elias2007Target}. In this approach, tandem mass spectra of peptides are searched against a database consisting of equal size of target and decoy protein sequences. The peptide-spectrum matches (PSMs) are scored and filtered by some score threshold. The FDR of selected PSMs is estimated by the ratio of the number of decoy matches to the number of target matches. Usually, the lowest score threshold is taken such that the estimated FDR is below a given level. Although this empirical target-decoy approach to FDR has been very effective in practice, its theoretical foundation was not established until we proved that a +1 correction to the number of decoy matches (the same as in equation \ref{STD_Con}) leads to rigorous FDR control under the assumption of independence between PSMs ~\citep{he2013Multiple}. Our work in the context of mass spectrometry was initially submitted to journals in 2013 (unpublished) and was made public in 2015 ~\citep{he2015theoretical}. The extension to general multiple testing as presented here was first described in an earlier manuscript ~\citep{he2018TDFDR}.

\citet{barber2015controlling} proposed the knockoff filter method for controlling the FDR when performing variable selection via Lasso regression for a Gaussian linear model. In this method, knockoff variables, which are not (conditionally on the original variables) associated with the response, are constructed and subjected to competition with the original variables (covariates). The basic rationale of knockoff filter in FDR control is identical to the target-decoy approach. First, knockoff is essentially synonymous with decoy in their roles. Second, the method used by knockoff filter to derive the rejection region, i.e., the FDR estimation formula with +1 correction and the procedure of selecting the score threshold, is exactly the same as the target-decoy approach. Third, after the proof of equal probabilities of a null variable obtaining a positive score (target label) or a negative score (decoy label), the proof of FDR control is the same mathematical problem addressed by the knockoff filter and the target-decoy approach, although their proving techniques are different. The main contribution of the knockoff filter is its sophisticated knockoff construction method that makes possible the proof of the aforementioned 'equal probabilities' for dependent variables. Knockoff filter allows the variables to be correlated with each other, but assumes the Gaussian noise in the linear model. In comparison, our approach (this paper) achieves FDR control for independent variables only, but puts no assumptions on the distribution of the variables. In addition, the original knockoff filter method required that the sample size ($n$) is no less than the number of variables ($p$) for FDR control.

\citet{candes2018panning} later re-framed the knockoff procedure and proposed the so-called model-X knockoffs method. Unlike the original linear model in which $X_{ij}$ was treated as fixed (stochasticity was from the Gaussian noise), the model-X knockoffs method treats $X_{ij}$ as random. It assumes knowledge of the joint distribution of the covariates, and constructs knockoffs probabilistically instead of geometrically. This removes the restriction on sample size ($n\ge{p}$) and makes the method applicable to both linear and non-linear models. Although the construction of model-X knockoffs does not rely on the specific distribution forms of the original variables in principle, Gaussian distribution is the only one that can be implemented at present. Another limitation of the knockoff method is its high computational cost on knockoff construction, which involves complex matrix computation, such as eigenvalue computation and semidefinite programming.

In the current knockoff methods, only one knockoff copy is constructed for each original variable, and the probability of a null variable or its knockoff copy being selected is equal (0.5). In our target-decoy procedure, multiple decoy permutations are constructed for each original variable, which offers us the flexibility of setting different probabilities of producing target or decoy tests for true null hypotheses.  This kind of multiple competition can enhance the power as we experimentally illustrated. Recently, \citet{Emery2019MultipleCompe} investigated the multiple competition problem in more depth. They presented two methods, namely max method and mirror method, for competition with the multiple decoys/knockoffs. The max method is most intuitive. It selects the variable (original or knockoff) with the highest score. \citet{Gimenez2019Multiple} also used the max method for multiple knockoffs. The mirror method is like what we do in our standard target-decoy procedure but is more flexible. It uses two adjustable rank cutoffs for target/decoy labelling, while we only use one adjustable cutoff for target labelling. \citet{Emery2019MultipleKnock} also proposed methods to construct multiple knockoffs that offer both FDR control and enhanced power. 

In recent years, the target-decoy/knockoff approach to FDR has attracted much attention from the field of statistics ~\citep{Srinivasan2020Compositional,Tian2020powerful,li2019GGM,jiang2020knockoff, katsevich2019Multilayer,Shen2019Cancer,Fan2019Stable, Liu2020ModelFree,Fan2020RANK,Romano2019Deep,Sesia2018Gene,barber2019highdim,barber2020Robust}. No doubt, this success was owed to the publication of the knockoff method by Candes et al. However, it should be noticed that we first proposed the FDR estimation formula with the +1 correction, which is the key to FDR control, and gave the first proof of FDR control (in the target-decoy framework under the independence assumption) ~\citep{he2013Multiple,he2015theoretical}. We also first introduced the multiple competition strategy ~\citep{he2018TDFDR}. These have been recognized by the community ~\citep{Levitsky2017unbiased,Keich2019Averaging,Danilova2019Bias,Coute2019Beyond,
Emery2019MultipleCompe,Emery2020thesis,Prieto2020Protein,Sulimov2020Tailor}


Other related works include that \citet{Levitsky2017unbiased} proposed an interpretation to the +1 correction based on the negative binomial distribution. However, this interpretation assumes that the number of null targets can be infinite and has uniform prior probability, and therefore, is not a rigorous interpretation.  ~\citet{storey2004strong} also had a +1 correction in their pFDR estimation to achieve FDR control. However, this correction was made to the number of $p$-values greater than a fixed threshold $\lambda$, which amounts to the total number of decoys in our case. This is very different from the target-decoy/knockoff approach in which the +1 correction is made to the number of decoys/knockoffs in the rejection region. 

\section{Conclusion}
\label{sec:Discussion}
In this paper, we presented the target-decoy approach to FDR control for multiple hypothesis testing. This approach is free of estimating the null distribution or the null proportion, and can rigorously control the FDR for independent variables. Simulation studies demonstrated its ability in FDR control and higher power than two representative traditional methods. 

In the target-decoy approach, the scores are only used to determine the labels and ranks of tests, and the statistical meaning of the scores is not required. Therefore, any test statistic can be used, regardless of whether or not its null distribution is known. This flexibility brings the potential to increase the power of multiple testing. In this paper, we only used the $t$-value. Trying other statistics or engineering specific scoring functions for different types of data is a topic worthy of future research. For example, machine learning-derived feature importance scores can in principle be directly used in our approach. 

In this paper, FDR control was proved for independent variables, and only simulation evaluation was performed for dependent variables. The theoretic analysis under dependency will be our future work. Especially, whether permutation-based decoys can lead to FDR control under some kind of dependency is an interesting problem that needs to be addressed. 

Moreover, our control theorem is based on the exchangeable hypothesis. This null hypothesis is stronger than the more popular hypothesis that the two groups have the same means. The performance of our approach for the `equality of means' hypothesis needs further studies. 

Finally, our approach can be extended to the pair-matched case-control study by adjusting Step 1 of the target-decoy procedure, i.e., randomly exchange the paired observed values just as the permutation tests for pair-matched study instead of permuting them. The other steps and analyses are the same. 

\vskip 14pt
\noindent {\large\bf Acknowledgements}

This work was supported by the National Key R\&D Program of China (2018YFB0704304) and the National Natural Science Foundation of China (32070668).
\par

\vskip 14pt
\noindent {\large\bf Supplementary Materials}

The supplementary material provides the proofs of theorems in the main text.
\par

\vskip 14pt
\noindent {\large\bf Software package}

The R package for the target-decoy procedure can be downloaded from \\http://fugroup.amss.ac.cn/software/TDFDR/package.zip.
\par
\vskip 14pt

\par

\markboth{\hfill{\footnotesize\rm Kun He, Mengjie Li, Yan Fu, Fuzhou Gong, Xiaoming Sun} \hfill}
{\hfill {\footnotesize\rm NULL-FREE FALSE DISCOVERY RATE CONTROL} \hfill}

\bibhang=1.7pc
\bibsep=2pt
\fontsize{9}{14pt plus.8pt minus .6pt}\selectfont
\renewcommand\bibname{\large \bf References}
\expandafter\ifx\csname
natexlab\endcsname\relax\def\natexlab#1{#1}\fi
\expandafter\ifx\csname url\endcsname\relax
  \def\url#1{\texttt{#1}}\fi
\expandafter\ifx\csname urlprefix\endcsname\relax\def\urlprefix{URL}\fi

\bibliographystyle{chicago}      
\bibliography{fdr}

\begin{thebibliography}{}

\bibitem[\protect\citeauthoryear{Almudevar, Klebanov, Qiu, Salzman, and
  Yakovlev}{Almudevar et~al.}{2006}]{almudevar2006utility}
Almudevar, A., L.~B. Klebanov, X.~Qiu, P.~Salzman, and A.~Y. Yakovlev (2006).
\newblock Utility of correlation measures in analysis of gene expression.
\newblock {\em NeuroRx\/}~{\em 3\/}(3), 384--395.

\bibitem[\protect\citeauthoryear{Barber and Cand{\`e}s}{Barber and
  Cand{\`e}s}{2015}]{barber2015controlling}
Barber, R.~F. and E.~J. Cand{\`e}s (2015).
\newblock Controlling the false discovery rate via knockoffs.
\newblock {\em The Annals of Statistics\/}~{\em 43\/}(5), 2055--2085.

\bibitem[\protect\citeauthoryear{Barber and Candès}{Barber and
  Candès}{2019}]{barber2019highdim}
Barber, R.~F. and E.~J. Candès (2019).
\newblock A knockoff filter for high-dimensional selective inference.
\newblock {\em Annals of Statistics\/}~{\em 47\/}(5), 2504--2537.

\bibitem[\protect\citeauthoryear{Barber, Candès, and Samworth}{Barber
  et~al.}{2020}]{barber2020Robust}
Barber, R.~F., E.~J. Candès, and R.~J. Samworth (2020).
\newblock Robust inference with knockoffs.
\newblock {\em Annals of Statistics\/}~{\em 48\/}(3), 1409--1431.

\bibitem[\protect\citeauthoryear{Basu, Cai, Das, and Sun}{Basu
  et~al.}{2018}]{basu2017weighted}
Basu, P., T.~T. Cai, K.~Das, and W.~Sun (2018).
\newblock Weighted false discovery rate control in large-scale multiple
  testing.
\newblock {\em Journal of the American Statistical Association\/}~{\em
  113\/}(523), 1172--1183.

\bibitem[\protect\citeauthoryear{Benjamini and Hochberg}{Benjamini and
  Hochberg}{}]{benjamini1995controlling}
Benjamini, Y. and Y.~Hochberg.
\newblock Controlling the false discovery rate: A practical and powerful
  approach to multiple testing.
\newblock {\em Journal of the Royal Statistical Society: Series B
  (Methodological)\/}~{\em 57\/}(1), 289--300.

\bibitem[\protect\citeauthoryear{Benjamini, Krieger, and Yekutieli}{Benjamini
  et~al.}{2006}]{benjamini2006adaptive}
Benjamini, Y., A.~M. Krieger, and D.~Yekutieli (2006).
\newblock Adaptive linear step-up procedures that control the false discovery
  rate.
\newblock {\em Biometrika\/}~{\em 93\/}(3), 491--507.

\bibitem[\protect\citeauthoryear{Benjamini and Yekutieli}{Benjamini and
  Yekutieli}{2001}]{benjamini2001control}
Benjamini, Y. and D.~Yekutieli (2001).
\newblock The control of the false discovery rate in multiple testing under
  dependency.
\newblock {\em Annals of statistics\/}~{\em 29\/}(4), 1165--1188.

\bibitem[\protect\citeauthoryear{Cand{\`e}s, Fan, Janson, and Lv}{Cand{\`e}s
  et~al.}{2018}]{candes2018panning}
Cand{\`e}s, E., Y.~Fan, L.~Janson, and J.~Lv (2018).
\newblock Panning for gold: model-x knockoffs for high dimensional controlled
  variable selection.
\newblock {\em Journal of the Royal Statistical Society: Series B (Statistical
  Methodology)\/}~{\em 80\/}(3), 551--577.

\bibitem[\protect\citeauthoryear{Chow and Teicher}{Chow and
  Teicher}{2012}]{chow2012probability}
Chow, Y.~S. and H.~Teicher (2012).
\newblock {\em Probability theory: independence, interchangeability,
  martingales}.
\newblock Springer Science \& Business Media.

\bibitem[\protect\citeauthoryear{Cout{\'e}, Bruley, and Burger}{Cout{\'e}
  et~al.}{2019}]{Coute2019Beyond}
Cout{\'e}, Y., C.~Bruley, and T.~Burger (2019).
\newblock Beyond target-decoy competition: stable validation of peptide and
  protein identifications in mass spectrometry-based discovery proteomics.
\newblock {\em bioRxiv:765057\/}.

\bibitem[\protect\citeauthoryear{Danilova, Voronkova, Sulimov, and
  Kertész-Farkas}{Danilova et~al.}{2019}]{Danilova2019Bias}
Danilova, Y., A.~Voronkova, P.~Sulimov, and A.~Kertész-Farkas (2019).
\newblock Bias in false discovery rate estimation in mass-spectrometry-based
  peptide identification.
\newblock {\em Journal of Proteome Research\/}~{\em 18\/}(5), 2354--2358.

\bibitem[\protect\citeauthoryear{Diz, Carvajal-Rodr{\'\i}guez, and
  Skibinski}{Diz et~al.}{2011}]{diz2011multiple}
Diz, A.~P., A.~Carvajal-Rodr{\'\i}guez, and D.~O. Skibinski (2011).
\newblock Multiple hypothesis testing in proteomics: a strategy for
  experimental work.
\newblock {\em Molecular \& Cellular Proteomics\/}~{\em 10\/}(3), M110--004374.

\bibitem[\protect\citeauthoryear{Efron}{Efron}{2004}]{efron2004large}
Efron, B. (2004).
\newblock Large-scale simultaneous hypothesis testing: the choice of a null
  hypothesis.
\newblock {\em Journal of the American Statistical Association\/}~{\em
  99\/}(465), 96--104.

\bibitem[\protect\citeauthoryear{Efron}{Efron}{2007}]{efron2007size}
Efron, B. (2007).
\newblock Size, power and false discovery rates.
\newblock {\em Annals of Statistics\/}~{\em 35\/}(4), 1351--1377.

\bibitem[\protect\citeauthoryear{Efron}{Efron}{2008}]{efron2008microarrays}
Efron, B. (2008).
\newblock Microarrays, empirical bayes and the two-groups model.
\newblock {\em Statistical Science\/}~{\em 23\/}(1), 1--22.

\bibitem[\protect\citeauthoryear{Efron}{Efron}{2012}]{efron2012large}
Efron, B. (2012).
\newblock {\em Large-scale inference: empirical Bayes methods for estimation,
  testing, and prediction}.
\newblock Cambridge University Press.

\bibitem[\protect\citeauthoryear{Efron and Tibshirani}{Efron and
  Tibshirani}{2002}]{efron2002empirical}
Efron, B. and R.~Tibshirani (2002).
\newblock Empirical bayes methods and false discovery rates for microarrays.
\newblock {\em Genetic epidemiology\/}~{\em 23\/}(1), 70--86.

\bibitem[\protect\citeauthoryear{Efron, Tibshirani, Storey, and Tusher}{Efron
  et~al.}{2001}]{efron2001empirical}
Efron, B., R.~Tibshirani, J.~D. Storey, and V.~Tusher (2001).
\newblock Empirical bayes analysis of a microarray experiment.
\newblock {\em Journal of the American statistical association\/}~{\em
  96\/}(456), 1151--1160.

\bibitem[\protect\citeauthoryear{Elias and Gygi}{Elias and
  Gygi}{2007}]{Elias2007Target}
Elias, J.~E. and S.~P. Gygi (2007).
\newblock Target-decoy search strategy for increased confidence in large-scale
  protein identifications by mass spectrometry.
\newblock {\em Nature Methods\/}~{\em 4\/}(3), 207--214.

\bibitem[\protect\citeauthoryear{Emery}{Emery}{2020}]{Emery2020thesis}
Emery, K. (2020).
\newblock {\em Controlling the FDR through multiple competition}.
\newblock Ph.\ D. thesis, The University of Sydney.

\bibitem[\protect\citeauthoryear{Emery, Hasam, Noble, and Keich}{Emery
  et~al.}{2019}]{Emery2019MultipleCompe}
Emery, K., S.~Hasam, W.~S. Noble, and U.~Keich (2019).
\newblock Multiple competition-based fdr control for peptide detection.
\newblock {\em arXiv:1907.01458\/}.

\bibitem[\protect\citeauthoryear{Emery and Keich}{Emery and
  Keich}{2019}]{Emery2019MultipleKnock}
Emery, K. and U.~Keich (2019).
\newblock Controlling the fdr in variable selection via multiple knockoffs.
\newblock {\em arXiv:1911.09442\/}.

\bibitem[\protect\citeauthoryear{Fan, Demirkaya, Li, and Lv}{Fan
  et~al.}{2020}]{Fan2020RANK}
Fan, Y., E.~Demirkaya, G.~Li, and J.~Lv (2020).
\newblock Rank: Large-scale inference with graphical nonlinear knockoffs.
\newblock {\em Journal of the American Statistical Association\/}~{\em
  115\/}(529), 362--379.
\newblock PMID: 32742045.

\bibitem[\protect\citeauthoryear{Fan, Lv, Sharifvaghefi, and Uematsu}{Fan
  et~al.}{2019}]{Fan2019Stable}
Fan, Y., J.~Lv, M.~Sharifvaghefi, and Y.~Uematsu (2019).
\newblock Ipad: Stable interpretable forecasting with knockoffs inference.
\newblock {\em Journal of the American Statistical Association\/}~{\em 0\/}(0),
  1--13.

\bibitem[\protect\citeauthoryear{Gimenez and Zou}{Gimenez and
  Zou}{2019}]{Gimenez2019Multiple}
Gimenez, J.~R. and J.~Zou (2019).
\newblock Improving the stability of the knockoff procedure: Multiple
  simultaneous knockoffs and entropy maximization.
\newblock Volume~89 of {\em Proceedings of Machine Learning Research}, pp.\
  2184--2192. PMLR.

\bibitem[\protect\citeauthoryear{He}{He}{2013}]{he2013Multiple}
He, K. (2013).
\newblock Multiple hypothesis testing methods for large-scale peptide
  identification in computational proteomics.
\newblock Master's thesis, University of Chinese Academy of Sciences.
\newblock
  \url{http://dpaper.las.ac.cn/Dpaper/detail/detailNew?paperID=20015738}.

\bibitem[\protect\citeauthoryear{He, Fu, Zeng, Luo, Chi, Liu, Qing, Sun, and
  He}{He et~al.}{2015}]{he2015theoretical}
He, K., Y.~Fu, W.-F. Zeng, L.~Luo, H.~Chi, C.~Liu, L.-Y. Qing, R.-X. Sun, and
  S.-M. He (2015).
\newblock A theoretical foundation of the target-decoy search strategy for
  false discovery rate control in proteomics.
\newblock {\em arXiv:1501.00537\/}.

\bibitem[\protect\citeauthoryear{He, Li, Fu, Gong, and Li}{He
  et~al.}{2018}]{he2018TDFDR}
He, K., M.~Li, Y.~Fu, F.~Gong, and X.~Li (2018).
\newblock A direct approach to false discovery rates by decoy permutations.
\newblock {\em arXiv:1804.08222\/}.

\bibitem[\protect\citeauthoryear{Jiang, Li, and Motsinger-Reif}{Jiang
  et~al.}{2020}]{jiang2020knockoff}
Jiang, T., Y.~Li, and A.~A. Motsinger-Reif (2020).
\newblock Knockoff boosted tree for model-free variable selection.
\newblock {\em arXiv:2002.09032\/}.

\bibitem[\protect\citeauthoryear{Katsevich and Sabatti}{Katsevich and
  Sabatti}{2019}]{katsevich2019Multilayer}
Katsevich, E. and C.~Sabatti (2019).
\newblock Multilayer knockoff filter: Controlled variable selection at multiple
  resolutions.
\newblock {\em Annals of Applied Statistics\/}~{\em 13\/}(1), 1--33.

\bibitem[\protect\citeauthoryear{Keich, Tamura, and Noble}{Keich
  et~al.}{2019}]{Keich2019Averaging}
Keich, U., K.~Tamura, and W.~S. Noble (2019).
\newblock Averaging strategy to reduce variability in target-decoy estimates of
  false discovery rate.
\newblock {\em Journal of proteome research\/}~{\em 18\/}(2), 585--593.

\bibitem[\protect\citeauthoryear{Kerr}{Kerr}{2009}]{kerr2009comments}
Kerr, K.~F. (2009).
\newblock Comments on the analysis of unbalanced microarray data.
\newblock {\em Bioinformatics\/}~{\em 25\/}(16), 2035--2041.

\bibitem[\protect\citeauthoryear{Langaas, Lindqvist, and Ferkingstad}{Langaas
  et~al.}{2005}]{langaas2005estimating}
Langaas, M., B.~H. Lindqvist, and E.~Ferkingstad (2005).
\newblock Estimating the proportion of true null hypotheses, with application
  to dna microarray data.
\newblock {\em Journal of the Royal Statistical Society: Series B (Statistical
  Methodology)\/}~{\em 67\/}(4), 555--572.

\bibitem[\protect\citeauthoryear{Lee, Efetova, Engelmann, Kramell, Wasternack,
  Ludwig-M{\"u}ller, Hedrich, and Deeken}{Lee
  et~al.}{2009}]{lee2009agrobacterium}
Lee, C.-W., M.~Efetova, J.~C. Engelmann, R.~Kramell, C.~Wasternack,
  J.~Ludwig-M{\"u}ller, R.~Hedrich, and R.~Deeken (2009).
\newblock Agrobacterium tumefaciens promotes tumor induction by modulating
  pathogen defense in arabidopsis thaliana.
\newblock {\em The Plant Cell\/}~{\em 21\/}(9), 2948--2962.

\bibitem[\protect\citeauthoryear{Levitsky, Ivanov, Lobas, and
  Gorshkov}{Levitsky et~al.}{2017}]{Levitsky2017unbiased}
Levitsky, L.~I., M.~V. Ivanov, A.~A. Lobas, and M.~V. Gorshkov (2017).
\newblock Unbiased false discovery rate estimation for shotgun proteomics based
  on the target-decoy approach.
\newblock {\em Journal of proteome research\/}~{\em 16\/}(2), 393--397.

\bibitem[\protect\citeauthoryear{Li and Maathuis}{Li and
  Maathuis}{2019}]{li2019GGM}
Li, J. and M.~H. Maathuis (2019).
\newblock Ggm knockoff filter: False discovery rate control for gaussian
  graphical models.
\newblock {\em arXiv:1908.11611\/}.

\bibitem[\protect\citeauthoryear{Liu, Ke, Liu, and Li}{Liu
  et~al.}{2020}]{Liu2020ModelFree}
Liu, W., Y.~Ke, J.~Liu, and R.~Li (2020).
\newblock Model-free feature screening and fdr control with knockoff features.
\newblock {\em Journal of the American Statistical Association\/}~{\em 0\/}(0),
  1--16.

\bibitem[\protect\citeauthoryear{Liu, Shao, et~al.}{Liu
  et~al.}{2014}]{liu2014phase}
Liu, W., Q.-M. Shao, et~al. (2014).
\newblock Phase transition and regularized bootstrap in large-scale $ t $-tests
  with false discovery rate control.
\newblock {\em The Annals of Statistics\/}~{\em 42\/}(5), 2003--2025.

\bibitem[\protect\citeauthoryear{Markitsis and Lai}{Markitsis and
  Lai}{2010}]{markitsis2010censored}
Markitsis, A. and Y.~Lai (2010).
\newblock A censored beta mixture model for the estimation of the proportion of
  non-differentially expressed genes.
\newblock {\em Bioinformatics\/}~{\em 26\/}(5), 640--646.

\bibitem[\protect\citeauthoryear{Meinshausen, Rice, et~al.}{Meinshausen
  et~al.}{2006}]{meinshausen2006estimating}
Meinshausen, N., J.~Rice, et~al. (2006).
\newblock Estimating the proportion of false null hypotheses among a large
  number of independently tested hypotheses.
\newblock {\em The Annals of Statistics\/}~{\em 34\/}(1), 373--393.

\bibitem[\protect\citeauthoryear{Prieto and Vázquez}{Prieto and
  Vázquez}{2020}]{Prieto2020Protein}
Prieto, G. and J.~Vázquez (2020).
\newblock Protein probability model for high-throughput protein identification
  by mass spectrometry-based proteomics.
\newblock {\em Journal of Proteome Research\/}~{\em 19\/}(3), 1285--1297.

\bibitem[\protect\citeauthoryear{Romano, Sesia, and Candès}{Romano
  et~al.}{2019}]{Romano2019Deep}
Romano, Y., M.~Sesia, and E.~Candès (2019).
\newblock Deep knockoffs.
\newblock {\em Journal of the American Statistical Association\/}~{\em 0\/}(0),
  1--12.

\bibitem[\protect\citeauthoryear{Sarkar}{Sarkar}{2002}]{sarkar2002some}
Sarkar, S.~K. (2002).
\newblock Some results on false discovery rate in stepwise multiple testing
  procedures.
\newblock {\em Annals of statistics\/}~{\em 30\/}(1), 239--257.

\bibitem[\protect\citeauthoryear{Scott and Berger}{Scott and
  Berger}{2010}]{scott2010bayes}
Scott, J.~G. and J.~O. Berger (2010).
\newblock Bayes and empirical-bayes multiplicity adjustment in the
  variable-selection problem.
\newblock {\em The Annals of Statistics\/}~{\em 38\/}(5), 2587--2619.

\bibitem[\protect\citeauthoryear{Sesia, Sabatti, and Candès}{Sesia
  et~al.}{2018}]{Sesia2018Gene}
Sesia, M., C.~Sabatti, and E.~J. Candès (2018).
\newblock Gene hunting with hidden markov model knockoffs.
\newblock {\em Biometrika\/}~{\em 106\/}(1), 1--18.

\bibitem[\protect\citeauthoryear{Shen, Fu, He, and Jiang}{Shen
  et~al.}{2019}]{Shen2019Cancer}
Shen, A., H.~Fu, K.~He, and H.~Jiang (2019).
\newblock False discovery rate control in cancer biomarker selection using
  knockoffs.
\newblock {\em Cancers\/}~{\em 11\/}(6).

\bibitem[\protect\citeauthoryear{Srinivasan, Xue, and Zhan}{Srinivasan
  et~al.}{2020}]{Srinivasan2020Compositional}
Srinivasan, A., L.~Xue, and X.~Zhan (2020).
\newblock Compositional knockoff filter for high-dimensional regression
  analysis of microbiome data.
\newblock {\em Biometrics\/}, biom.13336.

\bibitem[\protect\citeauthoryear{Storey}{Storey}{2002}]{storey2002direct}
Storey, J.~D. (2002).
\newblock A direct approach to false discovery rates.
\newblock {\em Journal of the Royal Statistical Society: Series B (Statistical
  Methodology)\/}~{\em 64\/}(3), 479--498.

\bibitem[\protect\citeauthoryear{Storey}{Storey}{2003}]{storey2003positive}
Storey, J.~D. (2003).
\newblock The positive false discovery rate: a bayesian interpretation and the
  q-value.
\newblock {\em The Annals of Statistics\/}~{\em 31\/}(6), 2013--2035.

\bibitem[\protect\citeauthoryear{Storey, Taylor, and Siegmund}{Storey
  et~al.}{2004}]{storey2004strong}
Storey, J.~D., J.~E. Taylor, and D.~Siegmund (2004).
\newblock Strong control, conservative point estimation and simultaneous
  conservative consistency of false discovery rates: a unified approach.
\newblock {\em Journal of the Royal Statistical Society: Series B (Statistical
  Methodology)\/}~{\em 66\/}(1), 187--205.

\bibitem[\protect\citeauthoryear{Storey and Tibshirani}{Storey and
  Tibshirani}{2003}]{storey2003statistical}
Storey, J.~D. and R.~Tibshirani (2003).
\newblock Statistical significance for genomewide studies.
\newblock {\em Proceedings of the National Academy of Sciences\/}~{\em
  100\/}(16), 9440--9445.

\bibitem[\protect\citeauthoryear{Strimmer}{Strimmer}{2008}]{strimmer2008unified}
Strimmer, K. (2008).
\newblock A unified approach to false discovery rate estimation.
\newblock {\em BMC bioinformatics\/}~{\em 9\/}(1), 303.

\bibitem[\protect\citeauthoryear{Sulimov and Kertész-Farkas}{Sulimov and
  Kertész-Farkas}{2020}]{Sulimov2020Tailor}
Sulimov, P. and A.~Kertész-Farkas (2020).
\newblock Tailor: A nonparametric and rapid score calibration method for
  database search-based peptide identification in shotgun proteomics.
\newblock {\em Journal of Proteome Research\/}~{\em 19\/}(4), 1481--1490.

\bibitem[\protect\citeauthoryear{Tan and Xu}{Tan and Xu}{2014}]{tan2014general}
Tan, Y.-D. and H.~Xu (2014).
\newblock A general method for accurate estimation of false discovery rates in
  identification of differentially expressed genes.
\newblock {\em Bioinformatics\/}~{\em 30\/}(14), 2018--2025.

\bibitem[\protect\citeauthoryear{Tian, Liang, and Li}{Tian
  et~al.}{2020}]{Tian2020powerful}
Tian, Z., K.~Liang, and P.~Li (2020).
\newblock A powerful procedure that controls the false discovery rate with
  directional information.
\newblock {\em Biometrics\/}, biom.13277.

\bibitem[\protect\citeauthoryear{Tusher, Tibshirani, and Chu}{Tusher
  et~al.}{2001}]{tusher2001significance}
Tusher, V.~G., R.~Tibshirani, and G.~Chu (2001).
\newblock Significance analysis of microarrays applied to the ionizing
  radiation response.
\newblock {\em Proceedings of the National Academy of Sciences\/}~{\em
  98\/}(9), 5116--5121.

\bibitem[\protect\citeauthoryear{Vergunst, van Lier, den Dulk-Ras, and
  Hooykaas}{Vergunst et~al.}{2003}]{vergunst2003recognition}
Vergunst, A.~C., M.~C. van Lier, A.~den Dulk-Ras, and P.~J. Hooykaas (2003).
\newblock Recognition of the agrobacterium tumefaciens vire2 translocation
  signal by the virb/d4 transport system does not require vire1.
\newblock {\em Plant physiology\/}~{\em 133\/}(3), 978--988.

\bibitem[\protect\citeauthoryear{Xie, Pan, and Khodursky}{Xie
  et~al.}{2005}]{xie2005note}
Xie, Y., W.~Pan, and A.~B. Khodursky (2005).
\newblock A note on using permutation-based false discovery rate estimates to
  compare different analysis methods for microarray data.
\newblock {\em Bioinformatics\/}~{\em 21\/}(23), 4280--4288.

\bibitem[\protect\citeauthoryear{Yu and Zelterman}{Yu and
  Zelterman}{2017}]{yu2017parametric}
Yu, C. and D.~Zelterman (2017).
\newblock A parametric model to estimate the proportion from true null using a
  distribution for p-values.
\newblock {\em Computational statistics \& data analysis\/}~{\em 114},
  105--118.

\end{thebibliography}


\end{document}